\documentstyle[aaspp4]{article}
\begin{document}


\title{Formation of Primordial Protostars}

\author{Kazuyuki Omukai and Ryoichi Nishi}
\affil{Department of Physics, Kyoto University, Kyoto 606-8502, Japan}


\begin{abstract}
 The evolution of collapsing metal free protostellar clouds is
 investigated for various masses and initial conditions. 
 We perform hydrodynamical calculations for spherically symmetric
 clouds taking account of radiative transfer of the molecular hydrogen
 lines and the 
 continuum, as well as of chemistry of the molecular hydrogen.    
 The collapse is found to proceed almost 
 self-similarly like Larson-Penston similarity solution. 
 In the course of the collapse, efficient three-body processes transform
 atomic hydrogen in an inner region of $\sim 1 M_{\sun}$ entirely into
 molecular form. 
 However, hydrogen in the outer part remains totally atomic although
 there is an intervening transitional layer of several solar masses,
 where hydrogen is in partially molecular form.  
 No opaque transient core is formed although clouds become optically
 thick to H$_{2}$ collision-induced absorption continuum, since H$_{2}$
 dissociation follows successively.  
 When the central part of the cloud reaches stellar densities ($\sim 
 10^{-2} {\rm g cm^{-3}}$), a very small hydrostatic core ($\sim
 5 \times 10^{-3} M_{\sun}$) is formed and subsequently grows in mass as the
 ambient gas accretes onto it.  
 The mass accretion rate is estimated to be $3.7 \times 10^{-2} M_{\sun}
 {\rm yr^{-1}} (M_{\ast}/M_{\sun})^{-0.37}$, where $M_{\ast}$ is 
 instantaneous mass of the central core, by using a similarity
 solution which reproduces the evolution of the cloud before the core
 formation.
\end{abstract}

\keywords{cosmology: theory --- early universe --- galaxies: formation
  --- stars: formation --- radiative transfer}

\newpage

\section{Introduction}
According to the standard scenarios of cosmology, the hydrogen
becomes neutral at $z\simeq 10^{3}$. 
The absence of Ly$\alpha$ trough in quasar spectra, however, severely
constrains the neutral fraction of the uniformly distributed intergalactic 
medium at $z<5$. 
This fact indicates that the hydrogen has to be reionized by energy 
injection from some objects in an intervening period. 
It is suggested that ultraviolet radiation from the known quasar
population is insufficient to photoionize the universe (Shapiro \&
Giroux 1987). 
Recently, the role of the first stars in reionizing the universe has
been emphasized by many authors (e.g., Fukugita \& Kawasaki
\markcite{FK94}1994; Ostriker \& Gnedin \markcite{OG96}1996; Haiman \&
Loeb \markcite{HL97}1997). 

However, star formation in the galaxy formation epoch is one of
the most poorly understood processes involved in galaxy formation.
Consequently, theories on the stellar reionization are undermined by many 
uncertainties, for example, the initial mass function, 
star formation rate and efficiency of the first stars.  
In order to resolve these uncertainties, detailed knowledge of the formation
processes of a first star from primordial gas seems to be required in
advance.
    
Primordial star formation can be viewed as the successive fragmentation and
contraction processes of collapsed cosmological objects. 
In this paper we discuss the latter process, namely the contraction of
primordial protostellar clouds \footnote{We use the term
  ``protostellar cloud'' to identify a gravitationally bound
  contracting gas cloud that, in part or as a whole, becomes a star as it 
  evolves. On the other hand, the term ``protostar'' means an accreting 
  core in hydrostatic equilibrium formed at the center of a protostellar
  cloud} into stars.
This problem has been investigated with a one-zone approximation by e.g.,
Carlberg (1981); Palla, Salpeter, \& Stahler (1983; see also Matsuda,
Sato, \& Takeda 1969; Yoneyama 1972).
   
On the other hand, Stahler, Palla, \& Salpeter \markcite{SPS86}(1986)
studied the main accretion phase of a primordial protostar assuming
stationary accretion of ambient material onto a central hydrostatic
core, which is a primordial counterpart of the present-day stellar core
(Larson\markcite{Lar69}1969). 
In that paper, they assumed Shu-like collapse (Shu
\markcite{Shu77}1977), i.e., the cloud initially in an unstable hydrostatic 
equilibrium is overrun by an expansion wave that propagates outward
from the center, leaving behind a growing central point mass and a
free-fall velocity field. 
Further, the mass accretion rate is constant with time and its rate is  \.{\it
  M}$ = 0.975 c_{\rm s}^{3}/G $, where $c_{\rm s}$ is the sound
velocity in the cloud.  
Higher temperatures, $\sim 10^3$ K, in primordial clouds results in 
a higher mass accretion rate $(4.41\times 10^{-3} M_{\sun} 
{\rm yr}^{-1})$ than the present-day value ($10^{-5} M_{\sun} {\rm
  yr}^{-1}$,    
e.g. Stahler, Shu, \& Taam \markcite{SST80}1980).
However, it remains unclear how the primordial protostellar cloud
reaches the main accretion phase.

Villere \& Bodenheimer \markcite{VB87}(1987) investigated this problem 
by using hydrodynamical methods.
They studied the spherically symmetric collapse of a primordial cloud of
$5 \times 10^{5} M_{\sun}$ into a protostar and pointed out the
importance of H$^{-}$ cooling for the determination of the minimum mass
of primordial stars. 
However, they did not treat the radiative transfer of H$_{2}$ lines, and
their efficiency of cooling in optically thick H$_{2}$ lines was
different from that of Palla et al. (1983).
Therefore, in order to follow the collapse of primordial protostellar
clouds accurately, we need to employ the hydrodynamical method together
with solving the radiative transfer of H$_{2}$ lines.
  
In this paper we study the evolution of primordial protostellar clouds by
performing hydrodynamical calculations, with radiative transfer of
H$_{2}$ lines and continuous radiation for spherically symmetric clouds 
up to the main accretion phase.

The outline of this paper is as follows. 
In \S 2, we describe the method of our calculation. 
In \S 3, results of calculations are presented. 
We summarize our work and discuss its implication in \S 4.
In the Appendix, our treatment of radiative transfer is described.

\section{Method of Calculation}
\subsection{Hydrodynamics}
We assume the cloud is spherically symmetric and neglect rotation, magnetic 
field, and external radiation field for simplicity.
The deviation from spherical symmetry and the effects of rotation may be 
important.
However, these effects make the problem two or three-dimensional and
intractable so far.     
In primordial star formation, magnetic fields are considered to play a
less important role than in present-day star formation.
Since star formation takes place in the deep interior of a protogalactic
cloud, and, moreover, in such a region the electron fraction is tiny, the
effects of cosmic background radiation are not significant.
 
The hydrodynamic code we use is a usual spherical symmetry,
explicit Lagrangian finite-difference scheme with
von Neumann-Richtmyer artificial viscosity (e.g., Mihalas \& Mihalas
1984; Thoul \& Weinberg 1995).

The Lagrangian equations of hydrodynamics describing the collapse of a
spherically symmetric cloud are 
the continuity equation
\begin{equation}
\frac{\partial m}{\partial r}=4 \pi r^2 \rho, 
\end{equation}
the momentum equation
\begin{equation}
\frac{Dv}{Dt}=-4 \pi r^2 \frac{\partial p}{\partial m}-\frac{Gm}{r^2},
\end{equation}
the energy equation
\begin{equation}
\frac{D \epsilon}{Dt}=-p\frac{D}{Dt}(\frac{1}{\rho})-\frac{\Lambda}{\rho}, 
\end{equation}
and the equation of state
\begin{equation}
p=(\gamma_{ad}-1)\rho \epsilon.
\end{equation}

In the above equations, the quantities $m,\rho,v,p,\varepsilon$, and $\Lambda$
are the mass within radius $r$, the density, velocity,
pressure, thermal energy per unit mass, and net cooling rate per unit
volume, respectively. 

The adiabatic exponent $\gamma_{\rm ad}$ is given by
\begin{equation}
\gamma_{\rm ad}=1+ \sum_{i=1}^{8} y_{i} /\sum_{i=1}^{8}
\frac{y_{i}}{\gamma_{i}-1},
\end{equation}
where the summation is over all chemical species, which include $i={\rm
  H^{0}}$,${\rm H^{+}}$,${\rm H^{-}}$,${\rm H_{2}}$,${\rm He^{0}}$,${\rm
  He^{+}}$,${\rm He^{++}}$, and $e^{-}$. 
The adiabatic exponent for H$_{2}$ is given by 
\begin{equation}
\label{eq;gamH2}
  \frac{1}{\gamma_{\rm H_{2}}-1}
=\frac{1}{2}[5+2x^{2}\frac{e^{x}}{(e^{x}-1)^{2}}]
, ~~x\equiv \frac{6100{\rm K}}{T}
\end{equation}
for H$_{2}$, where the second term in equation (\ref{eq;gamH2}) counts
the vibrational degrees of freedom and by $1/(\gamma_{i}-1)$=1.5 for
the other species.    
The concentration of the $i$ th species $y_{i}$ is defined by
\begin{equation}
y_{i}=\frac{n(i)}{n_{\rm H}},
\end{equation}
where $n(i)$ is the number density of the $i$ th species, and $n_{\rm H}=n({\rm
  H^{0}})+n({\rm H^{+}})+n({\rm H^{-}})+2n({\rm H_{2}})$ is 
the number density of hydrogen nuclei.
We do not include the radiation force term in the momentum equation (2),
since it is always negligible compared with the gas pressure in our runs. 
  
The net cooling rate $\Lambda$ consists of two parts, that is, the 
radiative cooling $\Lambda_{\rm{rad}}$ and the chemical cooling
$\Lambda_{\rm{chem}}$.  
The former can be written by using the luminosity $L(m)$ as
\begin{equation}
\Lambda_{\rm{rad}}(m)=\rho \frac{\partial L(m)}{\partial m}.
\end{equation} 
The luminosity $L(m)$ is obtained by solving radiative transfer as
described in \S 2.3 and the Appendix.  
We consider H$_{2}$ lines and continuum components that consist of
H$_{2}$ collision-induced absorption (CIA), H$^{-}$ bound-free absorption,
Lyman continuum absorption, etc., as sources of opacity. 
The net chemical cooling rate $\Lambda_{\rm{chem}}$ is given by
\begin{equation}
\Lambda_{\rm{chem}}(m)=-\rho \frac{\partial \epsilon_{\rm
    chem}}{\partial t},
\end{equation} 
where $\epsilon_{\rm chem}$ is the chemical binding energy per unit
mass.

We keep the external pressure constant with time as the boundary condition.
The timestep is the smaller of that determined by the Courant
condition (Courant number=0.2) and $10^{-3}$ of the local cooling time.  

\subsection{Chemical Reactions}
We consider the following eight species: ${\rm
  H^{0}}$,${\rm H^{+}}$,${\rm H^{-}}$,${\rm H_{2}}$,${\rm He^{0}}$,${\rm
  He^{+}}$,${\rm He^{++}}$, and $e^{-}$. 
The abundance of helium atoms is assumed to be 28 \% by mass. 

At low density ($n_{\rm H}< 10^{15} {\rm cm^{-3}}$), reaction equations
between ${\rm H^{0},H^{+},H^{-},H_{2}}$, and ${e^{-}}$ are 
solved using an implicit difference scheme. 
Considered reactions are the same as Palla et
al. (1983), while some reaction coefficients are
altered (Table 1).
The H$^{-}$ fraction is assumed to be the equilibrium value.
Helium atoms are considered to be all neutral, because in our case this low
density range corresponds to low temperature, namely $T<2100{\rm K}$.  

Before the number density exceeds $10^{15} {\rm cm^{-3}}$, the H$_{2}$
fraction easily reaches the equilibrium value. 
At $n_{\rm H}>10^{15} {\rm cm^{-3}}$ we switch the computation to the
equilibrium chemistry.   
Although the $e^{-}$ fraction has not reached equilibrium yet, this
does not cause any significant error, since $e^{-}$ plays no role in H$_{2}$
formation/dissociation at that time.    

The equilibrium chemistry ($n_{\rm H}>10^{15} {\rm cm^{-3}}$) is
computed by solving coupled Saha equations between
${\rm H^{0},H^{+},H_{2},He^{0},He^{+},He^{++}}$, and ${e^{-}}$.
 
\subsection{Radiative Transfer}
We consider H$_{2}$ rovibrational lines and continuum components as
sources of opacity and treat their transfer separately.  

H$_{2}$ line transfer is dealt with as a frequency
dependent transfer problem as is described in detail in the Appendix. 
We consider the first three vibrational 
states with rotational levels up to $J=20$ in each vibrational state
following Palla et al. (1983). 
The population of each level is determined by assuming local
thermodynamic equilibrium (LTE), since we consider the density range
well above the 
critical number density $n_{\rm{cr}}\sim 10^{4} {\rm cm^{-3}}$, where
radiative and collisional deexicitation rates become equal for the
molecular hydrogen.
We consider both thermal and kinematic Doppler shifts as a line broadening
mechanism.

Because of the lack of dust grains, primordial gas has only very weak
continuum opacity.  
Continuum transfer is treated as a gray problem using 
mean opacity. 
When the cloud becomes substantially optically thick, it becomes
difficult to keep numerical accuracy in solving the gray transfer
equation.
In such a case, we separate the cloud
into two regions, namely, the opaque interior and transparent envelope.
We solve the transfer equation only in the outer region, while the
diffusion approximation is employed in the interior.

As the mean opacity, we take the continuum-Planck mean opacity for
metal free gas at 1000K$<T<$7000K from Lenzuni, Chernoff \& Salpeter
(1991), which includes all the 
important continuum processes, specifically, bound-free absorption by
${\rm H^{0}}$ and ${\rm H^{-}}$, free-free absorption by ${\rm H^{0}}$,
${\rm H^{-}}$, ${\rm H_{2}}$, ${\rm H_{2}^{-}}$, ${\rm H_{2}^{+}}$,
${\rm H_{3}}$, ${\rm He^{0}}$, and ${\rm He^{-}}$, photodissociation of
${\rm H_{2}}$, and ${\rm H_{2}^{+}}$, Rayleigh scattering by ${\rm H^{0}}$,
${\rm H_{2}}$, and ${\rm He^{0}}$, Thomson scattering by $e^{-}$, and
collision-induced absorption by ${\rm H_{2}}$ due to collisions with
${\rm H_{2}}$, ${\rm He^{0}}$, and ${\rm H^{0}}$.
They omit all lines, for example, Ly$\alpha$ or ${\rm H_{2}}$
rovibrational (quadrupole) transitions in their opacity.
We treat the ${\rm H_{2}}$ lines as was described above. 
The Ly$\alpha$ lines are not so important in cooling in our case, because at
such high temperatures where they are excited, the density is so high and
their optical depth is so huge that these photons are always absorbed
before escaping the cloud.
At $T<1000{\rm K}$, we set the continuum opacity to be zero.
At $T>7000{\rm K}$, it is supplemented by using the Rosseland mean
opacity for metal free gas from Rogers \& Iglesias \markcite{RI92}(1992).
Our adoption of the Planck mean opacity at $T<7000{\rm K}$ and the
Rosseland mean opacity at higher temperature may deserve some comments. 
At low temperature, and then at low density,
protostellar clouds are transparent to continuum emission, so that Planck
mean opacity is appropriate. 
While the continuum optical depth exceeds unity $(T>2000{\rm K})$,
both mean opacities have close values within a factor of 2 in our
relevant range (Lenzuni et al. \markcite{LCS91}1991). 
Actually, they match smoothly to each other at the interface.

We solve H$_{2}$ line transfer and continuum transfer
independently, i.e., In solving H$_{2}$
line transfer, we neglect continuum opacity. 
However, errors caused by this are not significant.
The reason is as follows:
once the clouds become opaque to continuum opacity in a small
central region where all the evolution occurs essentially, H$_{2}$ line
cooling is negligible compared with continuum and chemical cooling there.
There still exists the outer region where H$_{2}$ line radiation is the
dominant cooling agent.   
However, in such a region the evolutionary timescale is so long in
comparison with the central one that this region remains practically
unchanged up to the formation of the hydrostatic core at the center (i.e.,
the end of our calculations).  
After the continuum optical depth exceeds unity, we set the cooling rate
due to H$_{2}$ lines equal to zero.
\section{Results}
\subsection{Initial Conditions}
We consider primordial protostellar clouds which are fragments of larger
ones.
Fragmentation of a collapsing cosmological object into primordial
protostellar clouds can be viewed as follows.
It is known that a spherical cloud in pressure-free collapse is unstable 
with respect to non-spherical perturbations (Lin, Mestel, \& Shu 1965;
Hutchins 1976; Susa, Uehara, \& Nishi 1996). 
As a consequence, the collapsing cosmological object first becomes a
pancaking disk, which is gravitationally unstable and fragments 
more easily into filamentary clouds rather than into spherical clouds
(Miyama, Narita, \& Hayashi 1987a, 1987b). 
Such a filamentary cloud is also gravitationally unstable and subject to
fragmentation but does not fragment immediately, as long as the
filamentary cloud collapses so fast that the density perturbation has
not sufficient time to grow (Inutsuka \& Miyama 1992, 1997).
Because the virial temperature of a filamentary cloud is
constant with time, the free-fall time $t_{\rm ff}$ decreases faster
than the cooling time $t_{\rm cool}$ as contraction proceeds.
Eventually, $t_{\rm ff} \simeq t_{\rm cool}$, i.e., the collapse becomes 
slow, then the filamentary cloud fragments. 

At the time of fragmentation, fragments cannot be highly 
gravitationally unstable. 
Hence, we take hydrostatic equilibrium
clouds with adiabatic stratification as the initial conditions for
simplicity.
Namely, the density distribution is represented by $\rho=\rho_{c}
(\theta_{n})^{n}$, where $\rho_{c}$ is the central density,
$n=1/(\gamma_{ad}-1)$ is the polytropic index, and $\theta_{n}$ is the
Emden function of index $n$.   
The pressure distribution is given by $p=K \rho^{\gamma_{ad}}$,
where $K$ is a free parameter fixed by giving the number density at
half mass radius $n_{h}$ and the total mass of the cloud $M$.
These clouds are cut off at the radius $r_{s}$ where the density falls
off $10^{-3}$ from the central value. 
Then parameters needed to specify an initial state are the mass of the
cloud $M$, the number density at the half mass radius $n_{h}$, and the
concentrations of chemical species.
The initial parameters for several models are given in Table 2.
Here, $N$ is the number of radial grids. 
As shown in Table 2, the initial width of the radial
zones is increased by constant ratio for each zone as distance from the
center is increased.  
This ratio is determined so as to resolve the smallest structure found in test
runs with a sufficient number of radial grids.
The mass and initial parameters of run A are close to those of a
primordial cylindrical cloud at the fragmentation epoch (Uehara 1998).
Hereafter, we call run A the fiducial run.
The parameters of run B are the same as those of run A, except the
number of radial grids is reduced to 100 in order to check the accuracy of our
calcualtion.
The match of the evolutionary paths between run A and run B is fairly good.
Therefore, radial grids of 100 cells are enough to follow the collapse as
long as the resolution around the center is sufficient.
Runs C and D correspond to smaller and denser fragments than the fiducial 
run, which result from the later fragmentation of cylindrical clouds
(Uehara et al. 1996). 
The evolution in these runs becomes almost the same as that of the
corresponding inner mass regions of the fiducial run.
If the H$_{2}$ fraction is too small at the begining, heat generated by
H$_{2}$ formation exceeds the radiative loss by H$_{2}$ lines and the cloud
expands until a sufficient amount of H$_{2}$ is produced.
However, this expansion is clearly a result of inappropriate initial
conditions.
We put as much H$_{2}$ fraction as necessary in these runs to avoid such
expansion.          
In the above runs, we set the H$_{2}$ concentration $f_{\rm H_{2}}={\rm
  const.}$ with radius initially, while in reality the $f_{\rm H_{2}}$
distribution has a gradient, that is, higher in the interior and lower in
the envelope.  
To see its effect, we set $f_{\rm H_{2}}=0$ initially in run E.
In this case the cloud does not collapse dynamically until a sufficient
amount of H$_{2}$ is formed.
It is found that, however, the evolution thereafter is the same as
in the fiducial case.  
      
As was mentioned above, the structure of collapsing primordial clouds becomes 
almost the same after initial transient phases that merely reflect
inappropriate initial conditions.    
Therefore, we describe the evolution in the fiducial run in detail hereafter.
  
\subsection{The collapse of a primordial protostellar cloud}
Initially the cloud is in hydrostatic equilibrium, and the cooling time
is somewhat longer than the free-fall time. 
For the fiducial run the cooling time $t_{\rm cool}\equiv \rho \epsilon /
\Lambda=5.1 \times 10^{5}$ yr and the free-fall time $t_{\rm
  ff}\equiv (3 \pi /32G \rho)^{1/2}=3.2 \times 10^{5}$ yr at the
center of the cloud initially.  
After some quasi-static contraction, the cooling time at the
center becomes shorter than the free-fall time.
At this time, the cloud begins to collapse dynamically in a free-fall
time scale.
In reality, the evolution should only be physically
relevant thereafter, since a fragment is in the state where $t_{\rm cool}\simeq
t_{\rm ff}$ at the fragmentation epoch (Low \& Lynden-Bell 1976; Rees
1976; Silk 1977; Uehara et al. 1996).

The evolutionary sequences of number density, temperature, velocity, and
H$_{2}$ concentration distributions are illustrated for the fiducial
run in Figure 1 and for run C, in Figure 3, as a function of 
the radial distance.    
As we can see from Figures 1 a, the collapse proceeds 
almost self-similarly and is very reminiscent of Larson-Penston's
similarity solution (Penston 1969; Larson 1969). 
Since the free-fall time depends inversely on the density, the collapse
proceeds most rapidly in the center where the density is the highest,
leaving the outer part practically unchanged.
The dash in the figure indicates the slope of density 
gradient for the isothermal case, namely $\rho \propto r^{-2}$
(Bodenheimer \& Sweigart 1968). 
The slope in our case is slightly steeper, and its value is about
$-2.2$. 
For the Larson-Penston similarity solution in the case of a polytropic
equation of state, i.e., $p=K \rho ^{\gamma}$, the density gradient may
be written as (Larson 1969)
\begin{equation} 
\frac{\partial {\rm ln} \rho}{\partial {\rm ln} r} = \frac{-2}{2- \gamma}. 
\end{equation} 
Therefore $\gamma \simeq 1.1$ in our case (see \S 3.3).

The evolution of the central temperature and the ratio of specific heats
$\Gamma=\frac{\partial {\rm ln} p/ \partial t}{\partial {\rm ln} \rho /
  \partial t}$ at the center are illustrated in Figure 3 a and 3 b.
According to Figure 3 b, the central value of $\Gamma$ is about 1.1 in
a large range of density, which coincides with the value expected from
the density gradient.   
An initial low value of $\Gamma$ at the center does not mean that the 
central part of the cloud cools efficiently at that time.
The reason is the following. 
In quasi-static contraction, although $\Gamma$ becomes nearly equal to the
critical value 4/3 on average over a whole cloud, this needs not be 
$\Gamma \simeq 4/3$ at the center.
Since the inner region cools faster than the outer part, a specific
entropy gradient (i.e., $\frac{\partial s}{\partial r}>0$) is
established in the course of contraction.
In the case that the cloud contracts from the initial adiabatic
stratification (i.e., $\frac{\partial s}{\partial r}=0$ ) to such a
state, $\Gamma$ tends to become lower in the inner region, while larger than
4/3 in the outer region in quasi-static contraction (see Omukai et
al. 1998). 
Our initial low value of $\Gamma$ at the center merely
reflects this fact. 
After the cloud starts dynamical collapse, efficient H$_{2}$-line cooling
makes the ratio of specific heats $\Gamma$ continue to decrease toward unity
(corresponding to the isothermal case) on average, until the cloud
becomes optically thick to several lines when the central number density
$n_{\rm c}\sim 10^{11} {\rm cm^{-3}}$.

Cooling due to molecular hydrogen formed by efficient three-body
processes, which become the dominant mode of H$_{2}$ production for
number density $n >10^{8} {\rm cm^{-3}}$, makes the temperature in the
central region even lower than in the outer part, as we can see in Figure 1
(b). 
These processes completely transform hydrogen in the central part of
about $1M_{\sun}$ into molecular form before the temperature becomes so
high that efficient H$_{2}$ dissociation begins.  
However, the outer region, i.e., $m>{\rm several}~M_{\sun}$ in mass
coordinate, remains totally atomic to the end (Figs. 1 and 2 d). 

By the time the central number density reaches about $10^{11} {\rm
  cm^{-3}}$ and the central temperature $T_{c}\simeq 900{\rm K}$, more
  than 60 \% of hydrogen atoms have been converted into molecular form in a
  small region around the center, and some of the H$_{2}$ lines begin to
  become optically thick.  
Eventually the cooling rate per unit mass around the
center becomes lower than that in the outer region. 
Although this raises the ratio of specific heats $\Gamma$
steadily, it does not exceed the critical value 4/3, as we can see in
Figure 3 (b).  
The reason is that there are always sufficient lines that are efficient in
cooling, namely, those whose optical depth is nearly unity.
   
When the central number density reaches about $3\times 10^{13} {\rm cm^{-3}}$
and the central temperature about $1600 {\rm K}$, the hydrogen molecules
begin to dissociate gradually. 
However, this is only temporary.
At the same time, H$_{2}$ CIA-continuum cooling comes into play. 
Continuum cooling is so strong as to stop and even to reverse 
dissociation, until it becomes optically thick and radiative cooling
is no longer efficient. 
After a small central part of the cloud becomes opaque ($ 3\times 10^{16}
{\rm cm^{-3}}, 2000 {\rm K} $), the full-scale dissociation begins
subsequently. 

In the case of the star formation in present-day molecular clouds, 
after the central region becomes opaque to dust continuum, a transient
core in hydrostatic equilibrium is formed with central density and 
temperature $n_{\rm c}=10^{14}{\rm cm^{-3}}$, and $T_{\rm c}=170 {\rm K}$
immediately after the formation (Larson 1969). 
In the primordial case, no transient core is formed when the central 
part of the cloud becomes optically thick to continuous radiation, since
H$_{2}$ dissociation follows successively. 
The density and temperature at the center at the begining of the full-scale
dissociation are nearly the same as in the present-day case.
Then the evolution also becomes nearly the same.

When most of the hydrogen molecules are dissociated, the ratio of
specific heats $\Gamma$ rises above the critical value 4/3 (Fig. 3 b). 
The partial ionization of the hydrogen atoms reduces $\Gamma$
temporarily but not 
below 4/3, since the density is already sufficiently high for
compressional work to dominate cooling owing to ionization.
After the central part of the cloud contracts almost adiabatically to
some extent, the hydrostatic 
core, whose mass is about $5 \times 10^{-3} {\rm M_{\sun}}$, forms at the
center when the central number density reaches $\sim 10^{22} {\rm
  cm^{-3}}$ and the central temperature $3 \times 10^{4} {\rm K}$. 
These physical dimensions are almost the same as those of the
stellar core found in the calculation of present-day star
formation. 
The formation of the core can be seen more clearly in the velocity
distributions (Figs. 1 c and 2 c).
When the central hydrostatic core forms, a shock front develops at the
surface.
The core grows in mass by accretion of envelope material and
eventually will become an ordinary star.

\subsection{The Mass Accretion Rate}
Because of our explicit method of hydrodynamics, the Courant condition makes
time steps extremely short and it becomes numerically very costly to
pursue the evolution after the core formation. 
Moreover our artificial viscosity method smears the accretion shock front, 
thus producing a small temperature gradient, which results in a low
luminosity and low radiative cooling when we use the diffusion approximation
(Winkler \& Newman \markcite{WN80}1980). 
We then stop our calculation at some arbitrary time after the core formation. 
Therefore the evolution thereafter cannot be known exactly. 

However, we found that the Larson-Penston-type similarity solution for
the equation of state $p=K \rho ^{\gamma}$ (Yahil \markcite{Yahil83}1983;
Suto \& Silk \markcite{SS88}1988), where $K=4.2\times 10^{11}$ (in 
cgs) and  $\gamma=1.09$ reproduces well our results of the evolution
in the fiducial run before the core formation (see Figure 4). 
Here we assume that the evolution of envelope after the central core formation
is also described by the same similarity solution.
Larson-Penston type similarity solutions can be extended naturally even
after the central density becomes infinite (i.e., core formation)
(Hunter \markcite{Hun77}1977; Yahil \markcite{Yahil83}1983).  
The solutions after the central density becomes infinite have the finite
point mass at the center (i.e., protostar), which grows in mass with
time as (Yahil \markcite{Yahil83}1983; Suto \& Silk \markcite{SS88}1988) 
\begin{eqnarray}
M_{\ast}&=&K^{3/2} G^{(1-3 \gamma)/2} t^{4-3 \gamma} m_{0} \nonumber \\ 
        &=&0.11 M_{\sun} (\frac{t}{1 {\rm yr}})^{0.73},
\label{eq:mass}
\end{eqnarray}
where $M_{\ast}$ is the mass of the central protostar, $t$ is the time
elapsed since the core formation, and $m_{0}$ is a 
non-dimensional constant whose value is 20 for $\gamma=1.09$
Larson-Penston type solution. 
We used $K=4.2\times 10^{11}$ and $\gamma=1.09$ in the second expression 
of equation (\ref{eq:mass}).
Thus the mass accretion rate to the central protostar is given by
\begin{eqnarray}
{\dot M_{\ast}}&=&(4-3 \gamma) K^{3/2} G^{(1-3 \gamma)/2}
t^{-3(\gamma-1)} m_{0} \nonumber \\
               &=&8.3 \times 10^{-2} M_{\sun} {\rm yr^{-1}} (\frac{t}{1
                 {\rm yr}})^{-0.27}.
\label{eq:acc}
\end{eqnarray}
Using the relation in equation (\ref{eq:mass}), we can write the mass accretion rate
as a function of an instantaneous mass of the protostar as
\begin{equation}
  {\dot M_{\ast}}=3.7 \times 10^{-2} M_{\sun} {\rm yr^{-1}}
  (\frac{M_{\ast}}{M_{\sun}})^{-0.37}.
\label{eq:acc2}
\end{equation}
According to equation (\ref{eq:acc}) or (\ref{eq:acc2}), the mass
accretion rate is huge and is diminishing with time. 
From the first expression of equation (\ref{eq:acc}), we can see that
there are three parameters that determine the mass accretion rate, i.e., 
$\gamma,K$ and $m_{0}$.
The decrease of the accretion rate with time is the common nature of
self-similar collapse with $\gamma>1$ as can be seen by the first
expression of equation (\ref{eq:acc}). 
The parameter $K$ is related to the temperature of protostellar clouds, and 
the parameter $m_{0}$ concerns the type of the similarity solution
(i.e., the Larson-Penston-type or the Shu-type solution, etc.) as
well as $\gamma$.   
Stahler et al. \markcite{SPS86}(1986) pointed out the higher mass accretion
rate $4.41 \times 10^{-3} M_{\sun} {\rm yr}^{-1}$ of primordial
protostars than present-day value because of higher temperature (then
the higher $K$ value) in primordial protostellar clouds, although they
thought of Shu-like collapse. 
In our analysis, however, the collapse proceeds like the Larson-Penston-type 
solution, whose parameter $m_{0}$ is about an order of magnitude higher
than that of the Shu-type solution.  
For example, in the isothermal ($\gamma=1$) Larson-Penston collapse the
density is 4.4 times higher than the hydrostatic equilibrium value and
the fluid velocity is 3.3 times than the sound speed at the time of
central core formation. 
This results in the value of $m_{0}$ and the mass accretion rate 48
times higher than that in isothermal Shu collapse of the same
temperature (Hunter \markcite{Hun77}1977).      
\section{Conclusion}
In this paper we have investigated the collapse of primordial clouds
into protostars using a spherically symmetric 
Lagrangian hydrodynamic code combined with radiative transfer of H$_{2}$ 
lines/continuum and chemical reactions.

The collapse proceeds self-similarly like the Larson-Penston
similarity solution until the central region reaches stellar
density.
During the collapse, the central part of about $1M_{\sun}$ becomes
fully molecular before the dissociation, while the outer part ($m> {\rm
  several}~M_{\sun}$ in mass coordinate) remains almost fully atomic.    
When the small central part of the cloud reaches stellar densities ($\sim
10^{-2} {\rm g~cm^{-3}}$), the hydrostatic core, whose physical
dimensions are roughly the same as that in present-day star formation,
is formed at the center of the cloud.
The mass accretion rate is typically $\sim
10^{-2} M_{\sun} {\rm yr^{-1}}$, which is about 3 orders of magnitude higher
than in the present-day case, and it declines with time.   

Our results suggest a stellar core of
$5 \times 10^{-3}M_{\sun}$ at its formation.
On the other hand, Palla et al. (1983) argued the minimum Jeans mass in
the primordial gas clouds to be $0.05M_{\sun}$ at H ionization. 
This mass can be identified with the mass of the stellar core at the time of
formation (Carlberg 1981).
This descrepancy seems to come not only from Palla et al. (1983)'s
one-zone treatment, but also from the fact that Palla et al. (1983) did
not take into account H$_{2}$ CIA continuum.  

\acknowledgements
We are grateful to H. Masunaga, H. Susa, and H. Uehara for fruitful
discussions, to H. Sato for his continuous encouragement,
to T. Chiba and N. Sugiyama for reading our manuscript carefully,
to S. Hayward for checking the English, and to the referee, F. Palla,
for improving this manuscript. 
We also would like to thank participants in the workshop ``Molecular
Hydrogen in the Early Universe'' (Firenze, 1997 December 4-6) for invaluable
comments.  
We owe thanks to the YITP computer system for the numerical analyses.
This work is supported in part by a Grant-in-Aid of Scientific Research
from the Ministry of Education, Science, Sports and Culture
No. 09740174(RN).

\newpage
\appendix
\section{Radiative Transfer in Spherical Symmetry}
In this Appendix, we describe our schemes 
of calculations of radiative transfer. 
The specific intensity $I_{\nu} ({\rm ergs}~{\rm sec^{-1}} 
{\rm cm^{-2}}{\rm sr^{-1}}{\rm Hz^{-1}})$ along a ray is calculated by
solving the radiative transfer equation (e.g., Rybicki \& Lightman
\markcite{RP}1979),
\begin{equation}
\frac{dI_{\nu}}{ds}=\alpha_{\nu}(S_{\nu}-I_{\nu}),
\label{eq:tr}
\end{equation}
where $s$ is the displacement along the ray, $\alpha_{\nu}$ is the
absorption coefficient, and $S_{\nu}$ is the source function.
  If the matter is in LTE,
$S_{\nu}=B_{\nu}(T)$.

In a dynamically collapsing cloud, supersonic fluid velocity develops,
then Doppler shift owing to bulk motion becomes comparable with thermal
Doppler width, and we have to take it into account.  
For H$_{2}$ lines, absorption coefficients can be written by using
Einstein $B$-coefficients,
\begin{equation}
\alpha_{\nu}=\frac{h \nu_{co}}{4 \pi}\phi (\nu_{co})
(n_{1}B_{12}-n_{2}B_{21}). 
\end{equation}
Here, $\nu_{co}=\nu (1-v_{t}/c)$ is the frequency in the comoving frame,
$v_{t}$ is the bulk velocity tangent to the ray, 
$n_{1}$ and $n_{2}$ are the number densities of molecules in the lower
and upper level of the transition, respectively, and $\phi(\nu)$ is the line
profile function. 
We only consider line broadening due to thermal Doppler effects. 
Then  
\begin{equation}
\phi(\nu)=\frac{1}{\Delta {\nu}_{\rm{D}} \sqrt{\pi}} e^{-(\nu-\nu_{0})^{2}
  /(\Delta {\nu}_{\rm{D}})^{2} }.
\end{equation}
Here, ${\nu}_{0}$ is the (comoving) line-center frequency and the
Doppler width $\Delta {\nu}_{\rm{D}}$ is defined by 
\begin{equation}
\Delta {\nu}_{\rm{D}}=\frac{{\nu}_{0}}{c}\sqrt{\frac{2kT}{\mu m_{\rm H}}}, 
\end{equation}
where $T$ is the temperature at each radius in the cloud, $m_{\rm H}$ 
is the mass of a hydrogen atom, and $\mu=2$ is the molecular 
weight for the molecular hydrogen.
We include the first three vibrational states, with rotational levels up 
to $j=20$ in each vibrational state following Palla et al. (1983). 
We use $A$-coefficients of molecular hydrogen from Turner, Kirby-Docken,
\& Dalgarno \markcite{TKD77}(1977). 
B-coefficients can be known from A-coefficients by the Einstein relations
\begin{equation}
g_{1}B_{12}=g_{2}B_{21},
\end{equation}
and 
\begin{equation}
A_{21}=\frac{2h {\nu_0}^{3}}{c^{2}} B_{21},
\end{equation}
where $g_{i}$ is the statistical weight of the level $i$. 
The critical number density $n_{\rm{cr}}$, where radiative
and collisional de-excitation rates become equal for the molecular hydrogen,
is $ \sim 10^{4} {\rm cm^{-3}}$. 
In our calculations, the number density is well above the critical
value, so that almost all of excited molecules would be deexcited by
collision with other molecules.  
Therefore, collisional LTE is established, and the level populations
are easily determined. 

Figure 5 shows the geometry of grids of radiative transfer calculation 
in the case of spherical symmetry (Hummer \& Rybicki
\markcite{HR71}1971; see also Masunaga, Miyama, \& Inutsuka
\markcite{MMI98}1998). 
To know the luminosity at the $i$-th radial 
mesh, $r_{i}$, we solve the transfer equation (\ref{eq:tr}) at a frequency
$\nu$ along from
the 0-th to $(i-1)$-th rays, and obtain the specific intensity at $r_{i}$ in
the direction $\theta_{ij}$, $I_{\nu}(r_{i},\theta_{ij}) ~(-i+1
\le j \le i-1)$. This procedure will be repeated for each
frequency mesh. We consider the frequency range for each line within
6 times the Doppler width corresponding to the central 
temperature from the line center of static media. 
Each line is separated into 30 frequency meshes in our
calculations, which is indicated to be sufficient by some
experiments. 
Integrating over $\theta$ and $\nu$, we obtain the
monochromatic energy flux $F_{\nu}$ and the luminosity $L$: 
\begin{equation}
F_{\nu}(r)=2 \pi \int_{0}^{\pi} I_{\nu}(r,\theta){\rm cos} \theta d\theta,
\end{equation}
\begin{equation}
L(r)=4\pi r^{2} \int_{0}^{\infty} F_{\nu}(r) d \nu .
\end{equation}
The cooling rate by lines whose optical depth at line center is less than 0.1
is calculated by using an optically thin approximation. 

We approximate continuum transfer as an grey transfer problem.
For grey material $\alpha_{\nu}=\alpha$, integrating over frequency we
have
\begin{equation}
\label{fitr}
\frac{dI}{ds}=\alpha(S-I),
\end{equation}
where
\begin{equation}
I \equiv \int_{0}^{\infty} I_{\nu}d \nu, 
S \equiv \int_{0}^{\infty} S_{\nu}d \nu.
\end{equation}
For LTE matter, $S=B(T)=\frac{ac}{4 \pi}T^{4}$.
We solve the frequency integrated transfer equation (\ref{fitr}) along
the same rays as in Figure 5, and integrating over the polar angle
$\theta$ we obtain the continuum luminosity.  

As the cloud contracts, the increment of optical depth between adjacent
radial grids $\Delta \tau$ around the center exceeds unity. 
In such a case, we can not resolve the temperature gradient precisely, and the 
numerical accuracy may be severely damaged in solving the gray transfer
equation. 
Then we treat the inner opaque region and the outer transparent
region separately. 
In the inner region the diffusion equation is solved, i.e., the
continuum energy flux is given by 
\begin{equation}
F_{\rm cont}=-\frac{16 \sigma T^{3}}{3\alpha} \frac{\partial T}{\partial r},
\end{equation}
and in the outer region the transfer equation (\ref{fitr}) is solved as
described above.  
These solutions for continuum energy flux are matched at the
interface, which is chosen so as to satisfy the following conditions (Narita,
Nakano, \& Hayashi \markcite{NNH70} 1970):
\begin{equation}
|\frac{B(r)-I(r,\pi)}{I(r,\pi)}|<0.2,
\end{equation}
\begin{equation}
\tau(r)>4,
\end{equation}
and 
\begin{equation}
\Delta \tau>1,
\end{equation}
where $I(r,\pi)$ is the inward intensity, and $\tau$ is the optical
depth that is measured from the surface in the radial direction.
Under these conditions, the error in $F_{\rm cont}(r)$ is less than 4
percent.


\newpage

\begin{deluxetable}{cccr}
\tablecaption{Chemical Reactions
\label{table1} }
\tablehead{
\colhead{number} &
\colhead{reaction}    & \colhead{reference} & 
}
\startdata
1& ${\rm H^{+}}+e^{-} \rightarrow {\rm H}+\gamma $ 
 & Abel et al. 1997, Janev et al. 1987 \tablenotemark{a}
\nl
2& ${\rm H}+e^{-} \rightarrow {\rm H^{-}}+\gamma $
 & Abel et al. 1997 \nl
3& ${\rm H}+{\rm H^{-}} \rightarrow {\rm H_{2}}+e^{-}$
 & Abel et al. 1997 \nl
4,5& ${\rm 3H} \rightleftharpoons {\rm H_{2}}+{\rm H}$ 
 & Palla et al. 1983 \nl 
6,7& ${\rm 2H}+{\rm H_{2}} \rightleftharpoons {\rm 2H_{2}}$
 & Palla et al. 1983 \nl 
8& ${\rm H}+e^{-} \rightarrow {\rm H^{+}}+2e^{-}$
 & Abel et al. 1997 \nl
9& ${\rm 2H}\rightarrow {\rm H^{+}}+{\rm H}+e^{-}$
 & Palla et al. 1983 \nl
\enddata
\tablenotetext{a}{ The recombination
   rate to the ground level (Janev et al. 1987) is subtracted from the
   case A recombination rate (Abel et al. 1997).}
\end{deluxetable}

\newpage

\begin{deluxetable}{cccccccr}
\tablecaption{Calculated Parameters
\label{table2} }
\tablehead{
\colhead{run} &
\colhead{$M/M_{\sun}$}  & \colhead{$n_{\rm h}({\rm cm^{-3}})$} &
\colhead{$y_{\rm H_{2}}$}  & \colhead{$y_{e^{-}}$} & 
\colhead{$N$} & \colhead{$\Delta r_{i+1}/ \Delta r_{i}$} &
}
\startdata
A (fiducial)& $10^{2}$   &   $10^{6}$   &   $5 \times 10^{-4}$   
 & $10^{-10}$  & 250 & 1.01 \nl
B& $10^{2}$   &   $10^{6}$   &   $5 \times 10^{-4}$   
 & $10^{-10}$  & 100 & 1.05 \nl
C& $10^{1}$   &   $10^{8}$   &   $5 \times 10^{-3}$   
 & $10^{-10}$  & 100 & 1.03 \nl
D& $10^{0}$   &   $10^{10}$   &   $4.5 \times 10^{-1}$   
 & $10^{-10}$  & 100 & 1.01 \nl
E& $10^{2}$   &   $10^{6}$   &   0   
 & $10^{-8}$  & 100 & 1.05 \nl
\enddata
\end{deluxetable}


\newpage

\begin{figure}
\plotone{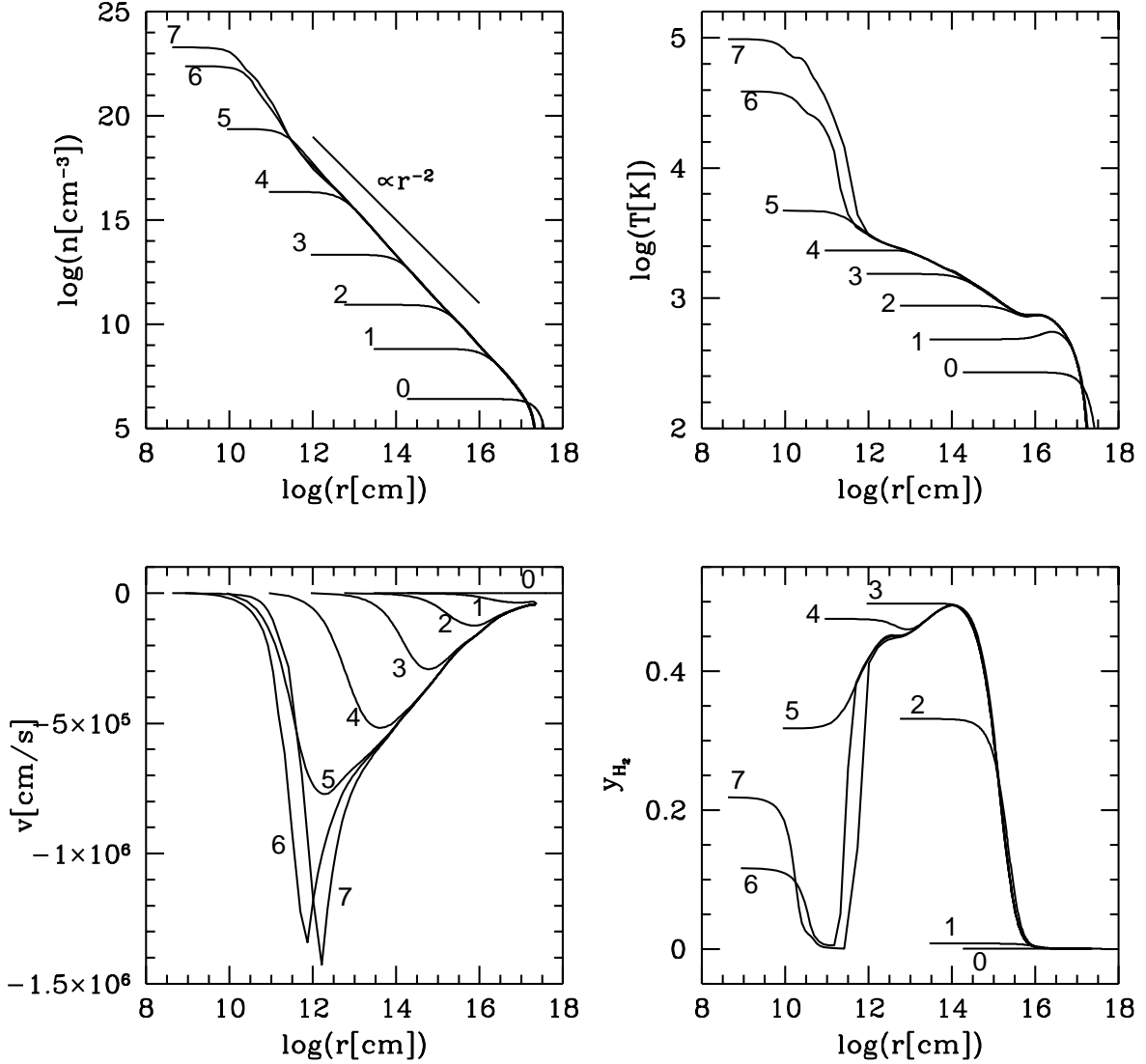}
\caption{The evolutionary sequences for the fiducial run. (a) The number 
  density ($n\equiv \rho/m_{\rm H}$, where $m_{\rm H}$ is the mass of the
  hydrogen atom), (b) temperature, (c) velocity, and (d) the
  H$_{2}$ concentration distribution versus the radial distance. 
  (0) initial state of our calculation: 
  (1) $5.7 \times 10^{5}$ yr after (0). 
  Three-body processes are active in the central region and temperature
  inversion occurs: 
  (2) $8.5 \times 10^{3}$ yr after (1).
  The cloud becomes optically thick to some lines:
  (3)  $2.8 \times 10^{2}$ yr after (2).
  The central region becomes fully molecular:
  (4) 12 yr after the state of (3).
  The central region becomes optically thick to H$_{2}$ CIA continuum:
  (5) 0.32 yr after (4).
  In the midst of dissociation: 
  (6) $2.4 \times 10^{-2}$ yr after (5).
  Shortly after the core formation:
  (7) $4.1 \times 10^{-2}$ yr after (6).
  Final state of our calculation.
\label{fig1}}
\end{figure}

\begin{figure}
\plotone{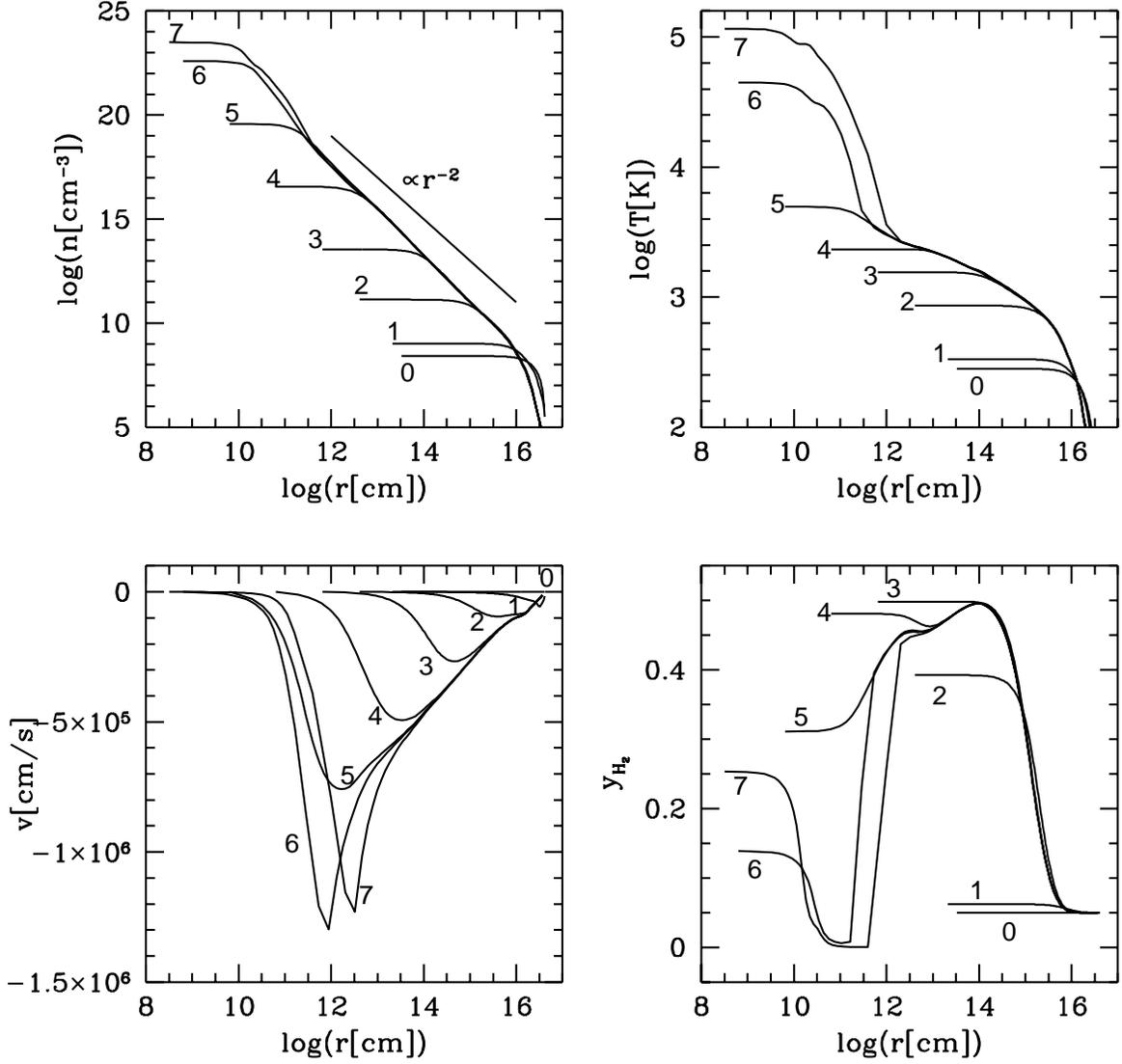}
\caption{The same as Figure 1 for run C. 
  (0) initial state of our calculation: 
  (1) $1.1 \times 10^{4}$ yr after (0). 
  Three-body processes are active in the central region: 
  (2) $7.8 \times 10^{3}$ yr after (1).
  The cloud becomes optically thick to some lines:
  (3)  $2.6 \times 10^{2}$ yr after (2).
  The central region becomes fully molecular:
  (4) 11 yr after (3).
  The central region becomes optically thick to H$_{2}$ CIA continuum:
  (5) 0.26 yr after (4).
  In the midst of dissociation: 
  (6) $2.5 \times 10^{-2}$ yr after (5).
  Shortly after the core formation:
  (7) $9.1 \times 10^{-2}$ yrs after (6).
  Final state of our calculation.
\label{fig2}}
\end{figure}

\begin{figure}
\plotone{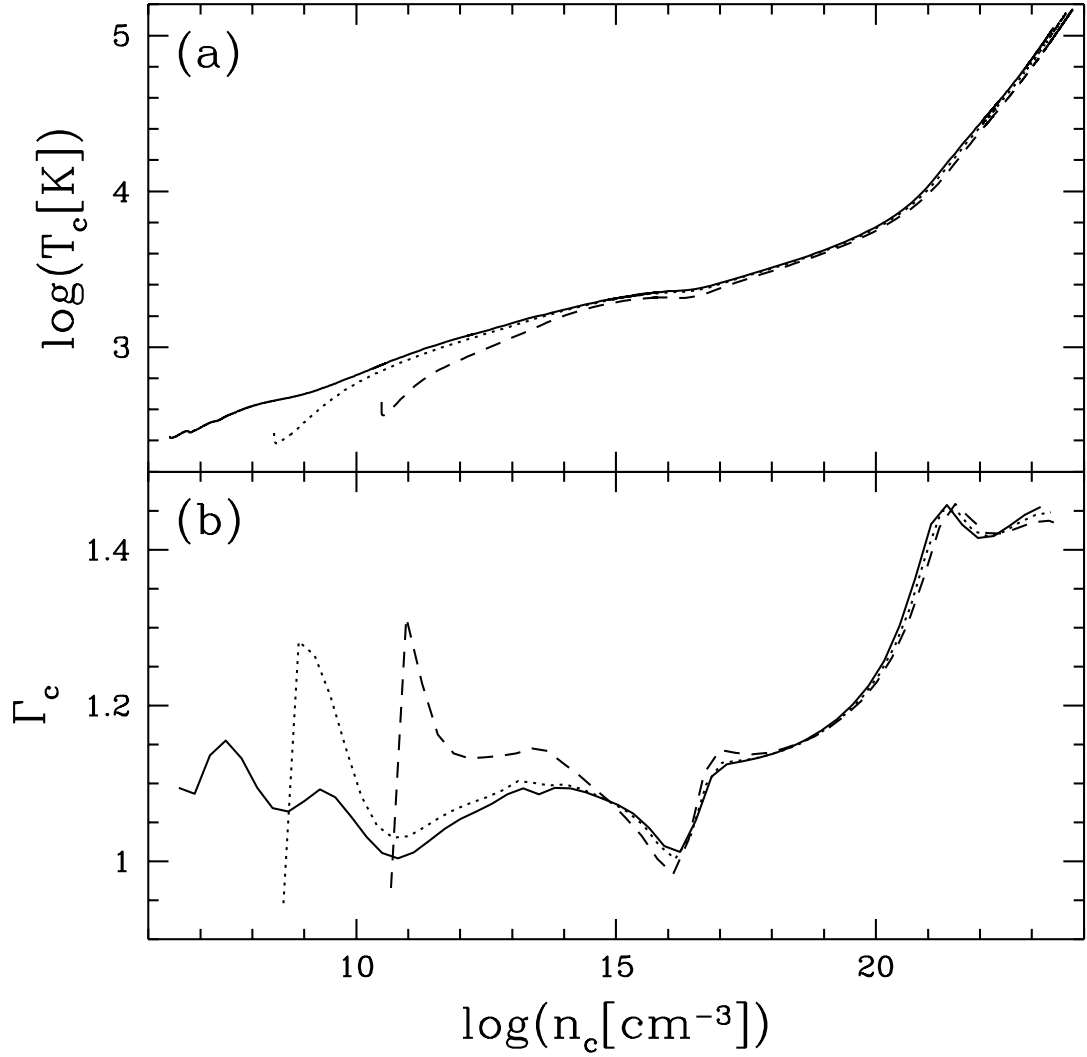}
\caption{The central evolution in runs A, C, and D. 
  (a) The central
  temperature $T_{c}(\rm K)$ is plotted over the central number density
  $n_{c} ({\rm cm^{-3}})$.
  (b) The evolution of the ratio of specific heats $\Gamma=\frac{\partial
  {\rm ln}p/ \partial t} {\partial {\rm ln}\rho/ \partial t}$ at
  the center. 
  In both panels the solid lines, the dotted lines, and the dashed lines
  represent runs A, C, and D, respectively.
\label{fig3}}
\end{figure}

\begin{figure}
\plotone{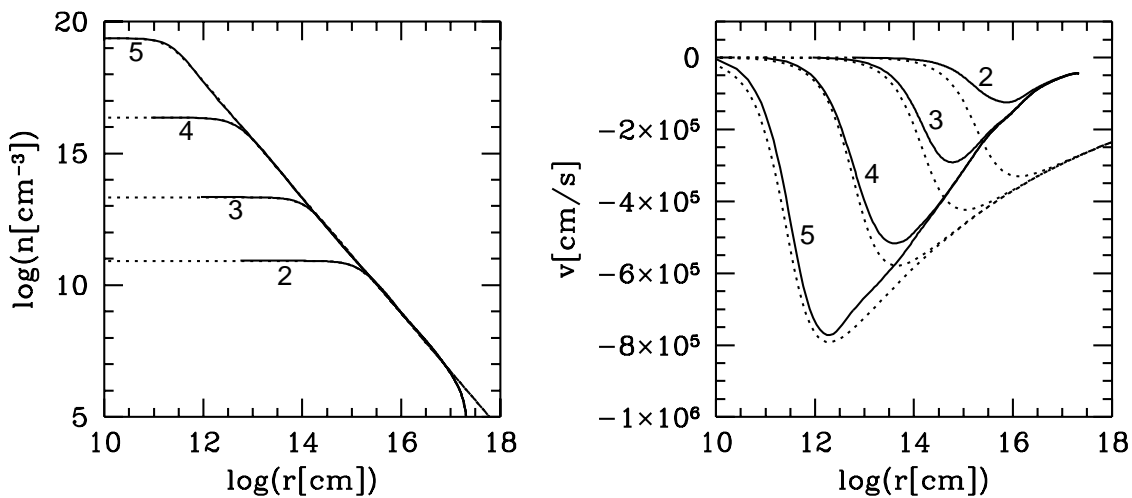}
\caption{Comparison between the evolution in fiducial run and the
  Larson-Penston type similarity solution.  
  Solid lines show (a)the number density and (b)the velocity
  distribution for the state 
  of 2,3,4, and 5 in the fiducial run (see Figure 1). 
  Dotted lines indicate the Larson-Penston type similarity solution for
  the equation of state $p=K \rho ^{\gamma}$, where $K=4.2\times 10^{11}$
  (in cgs) and $\gamma=1.09$, at the time $0.96 \times 10^{-2}, 0.31,
  10$, and $1.6 \times 10^{2}$ yrs before core formation. 
  We choose these parameters to fit the density distribution. 
  Although normalization of velocity is smaller and therefore
  timescale of contraction is longer in our calculation than in the
  similarity solution due to the influence of initial and boundary
  conditions, they also shows convergence to those of the similarity
  solution.
\label{fig4}}
\end{figure}

\begin{figure}
\plotone{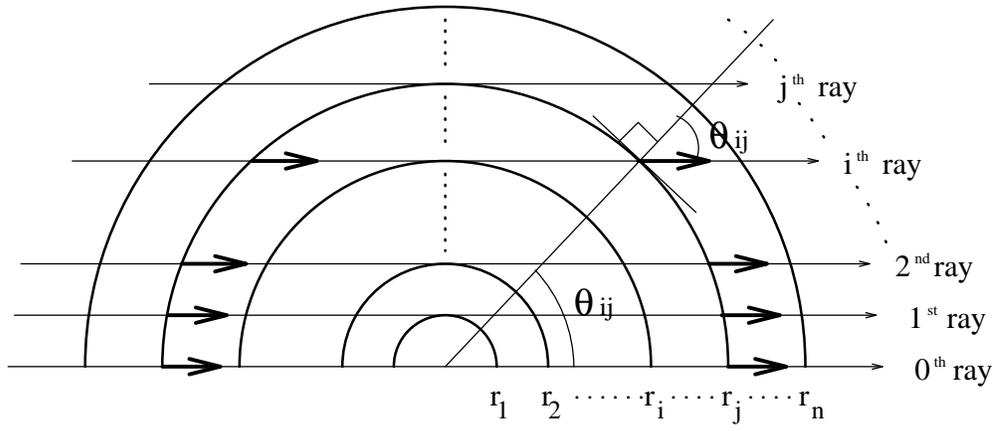}
\caption{Ray geometry of radiative transfer calculation in spherical
  symmetry. We solve transfer equation along rays tangent to each mass
  shell and obtain the intensity $I_{\nu}(\theta)$ at each grid point.
\label{fig5}}
\end{figure}

\end{document}